\documentclass[final,1p,times]{elsarticle}

\usepackage{amssymb}

\usepackage{lineno}

\usepackage{xcolor}
\usepackage{comment}
\usepackage{braket}
\usepackage{mathtools}
\usepackage{gensymb}
\usepackage[utf8]{inputenc}

\journal{Nuclear Instruments and Methods in Physics Research Section A}

\begin{document}

\begin{frontmatter}



\title{Characterization of newly developed large area SiC sensors for the NUMEN experiment}

\author[lns]{D. Carbone}
\author[lns]{A. Spatafora}
\author[to]{D. Calvo}
\author[unito]{F. Guerra}
\author[lns]{G.A. Brischetto}
\author[lns,unict]{F. Cappuzzello}
\author[lns]{M. Cavallaro}
\author[to]{M. Ferrero}
\author[imm]{F. La Via}
\author[lns]{S. Tudisco}
\author[]{for the NUMEN collaboration}

\affiliation[lns]{organization={INFN-Laboratori Nazionali del Sud},
             city={Catania},
             country={Italy}}
\affiliation[to]{organization={INFN-Sezione di Torino},
             city={Torino},
             country={Italy}}
\affiliation[unito]{organization={Università degli Studi di Torino, Dipartimento di Fisica},
             city={Torino},
            country={Italy}}
\affiliation[unict]{organization={Dipartimento di Fisica e Astronomia "Ettore Maiorana", Università di Catania},
             city={Catania},
             country={Italy}}
\affiliation[imm]{organization={ Institute for Microelectronics and Microsystems (IMM), National Research Council (CNR)},
             city={Catania},
             country={Italy}}


\begin{abstract}
First prototypes of large area, p-n junction, silicon carbide (SiC) detectors have been produced as part of an ongoing programme to develop a new particle identification wall for the focal plane detector of the MAGNEX magnetic spectrometer, in preparation for future NUMEN experimental campaigns. First characterizations of sensors from two wafers obtained with epitaxial silicon carbide growth and with different doping concentration are presented. Current (I-V) and capacitance  (C-V) characteristics are investigated in order to determine the full depletion voltage and the doping profile. Radioactive $\alpha$-sources are used to measure the energy resolution and estimate the depletion depth.  
\end{abstract}







\end{frontmatter}


\section{Introduction}
In recent years, there has been a growing interest in using Silicon Carbide (SiC) detectors for harsh environments. In particular, they are becoming increasingly popular in nuclear physics experiments due to their excellent ability to operate under high incident ions fluxes \cite{he2024,altana2023,Ciampi2019,raciti2010,ruddy2006}. In the case of small SiC detectors (5 × 5 mm$^2$, 10 $\mu$m thick), irradiated with $^{16}$O and $^{27}$Al ions, it was proven that such detectors could tolerate ﬂuencies as large as 10$^{13}$ heavy ions/cm$^2$, for the case where the heavy ions are stopped inside the detectors \cite{altana2023}.
In addition, SiC detectors have recently benefited from technological improvements resulting from the SiCILIA project \cite{tudisco2018}, that established new production processes able to build large area SiC devices with epitaxial layers as thick as 100 $\mu$m. 

Radiation hardness in detectors is a fundamental requirement for experiments where very low cross sections of rare processes have to be measured amidst high levels of background originating from more probable processes. Indeed, in order to maximize the yield of the rare processes under investigation, the challenge is to reduce the background to an acceptable level. This is the case for the NUMEN experiment \cite{numen2018,cappuzzello2021TDR}, which requires accurate cross section measurements of heavy-ion induced Double Charge-Exchange (DCE) reactions, aiming to access experimentally driven information on nuclear matrix elements of neutrinoless double beta decay (0$\upsilon\beta\beta$) \cite{Agostini2023,cappuzzello2023}. The DCE ($^{18}$O,$^{18}$Ne) and ($^{20}$Ne,$^{20}$O) reactions are explored at the Laboratori Nazionali del Sud of the Istituto Nazionale di Fisica Nucleare (INFN-LNS) in Catania using the Superconducting Cyclotron to accelerate the beams, and the MAGNEX large acceptance magnetic spectrometer \cite{magnex_review,cavallaro2020} to detect the reaction products. Since the DCE cross sections are quite low (tens of nb \cite{soukeras2021}, which is more than 8 orders of magnitude less than the other competing reaction channels from the same collision), a major upgrade of the LNS facility was launched to increase the beam intensities up to $10^{13}$ pps. To this purpose, technology at the forefront of scientific discovery are being developed for the accelerator, the target system, as well as the detection systems coupled to the MAGNEX spectrometer \cite{ciraldo2023,calvo2022,agodi2021,finocchiaro2020,iazzi2017}. In particular, the focal plane detector of MAGNEX will be equipped with a new particle identification (PID) system, composed of $\Delta$E-E telescopes. Large area SiC detectors will be used as the $\Delta$E stage and Cesium Iodide (Tl doped) crystals will detect the residual energy. The new MAGNEX PID system must guarantee an unprecedented hardness to heavy ions, neutrons, and gamma rays while at the same time adhere to stringent energy and time resolutions as well as efficiency requirements. For these reasons SiC sensors are considered to be the best choice as the $\Delta$E detectors. 

To be relevant for the NUMEN project the thickness of the SiC detectors has to allow for the detection of the ejectiles of interest (4 $\leq$ Z $\leq$ 10) over a wide dynamical range of incident energies (10 $\leq E/A\le$ 60 AMeV). An appropriate thickness for the $\Delta$E SiC stage is $\approx$ 100 $\mu$m \cite{cappuzzello2021TDR}. Since 720 SiC detectors will be used to cover the full MAGNEX detection area (about 0.2 m$^2$), homogeneity among different detectors in terms of depletion voltage, thickness, charge collection efficiency, and resolution is needed. Another crucial aspect for the MAGNEX PID system is that it will work in a low pressure gas environment of the new gas tracker (typically tens of mbar of isobutane or other gases) \cite{ciraldo2023,finocchiaro2020}, thus the high voltage needed to fully deplete the SiC detectors should be as low as possible, to avoid discharges propagating in the gas and causing damage and instabilities in the detector response. This NUMEN specific requirement therefore demands a decrease of the doping concentration to less than 10$^{14}$ atm/cm$^3$ to reduce the full depletion voltage to few hundreds of volts. 

In this work we present several characterizations of state-of-the-art large area SiC detectors that were recently produced to satisfy the requirement of the NUMEN project. To our knowledge, this is the first time that SiC detectors with such large area ($\approx$ 2.4 cm$^2$) and thick epitaxial layer ($\approx$ 100 $\mu$m) are produced and characterized. In particular, we measured their current (I--V) and capacitance (C--V) characteristics curves, in order to determine the full depletion voltage (FDV) of each sensor, and to study the doping profile. We also irradiated the sensors with radioactive $\alpha$-sources, to measure their energy resolution, linearity, and dead layer thickness. 

The paper is organized as follows. First, we describe the main properties of SiC in Section \ref{sec:sic_properties}, followed by a detailed discussion of the prototype sensors in Section \ref{sec:prototype_sensors}. Section \ref{sec:iv_cv} is devoted to the description of the I--V and C--V characteristics, and Section \ref{sec:doping} provides details of the doping profile. Measurements with the radioactive sources are discussed in Section \ref{sec:alpha}, and final conclusions are provided in Section \ref{sec:conclusions}.

\section{Silicon Carbide properties}
\label{sec:sic_properties}

SiC is a compound semiconductor with a wide bandgap ($\approx$ 3.28 eV). Among the many different polytypes of SiC (more than 200), the 4H-SiC polytype is considered the most appropriate for particle detector applications \cite{Hazdra2019,Hazdra2021,nava2008}, and is characterized by an energy bandgap of 3.23 eV. 
An important characteristic of SiC detectors is the radiation hardness, i.e. the resistance to damage caused by high doses of particle irradiation, where the traversing particles damages the lattice mainly by displacing a primary knock-on atom out of its lattice site. For 4H-SiC detectors this process has a much larger energy threshold ($\approx$ 25 eV) compared to the case of silicon (Si) detectors ($\approx$ 15 eV), which makes SiC a good alternative to Si.

The wide bandgap of SiC is also useful, as it significantly reduces the rate of thermal noise. SiC devices also present much smaller leakage current with respect to Si detectors \cite{Sze2006}. In fact, the reverse current density of SiC p-n junction is expected to be about 3 orders of magnitude smaller than for Si. On the other hand, a particle with a certain energy, ideally converting all its energy to the production of electron-hole pairs, generates about 3 times more charge carriers in Si (bandgap 1.12 eV) than in SiC. Detectors based on SiC, therefore, feature lower pulse amplitudes. However, if we consider applications for the heavy-ion detection, as in the case of the NUMEN project, many primary charge carriers are generated, whose statistical fluctuations are not an issue. In addition, SiC detectors maintain a good signal-to-noise ratio at temperatures at which Si detectors require external cooling to keep the intrinsic carrier level sufficiently low~\cite{Lebedev2004}.

For time resolution, SiC detectors can profit from the high saturation velocities of the charge carriers (2 x 10$^7$ cm/s), which is twice that in Si. Also it is possible to operate the devices at or close to the carrier velocity saturation condition. This is because the breakdown field in SiC is 3 MV/cm, ten times higher than in Si. The junctions on SiC can hence reach an extremely high internal electric field in the depleted region. Values for the electric field as high as 10$^5$ V/cm has been reached without any junction breakdown or significant increase of the reverse current \cite{lioliou2016}. A timing resolution of hundreds of ps has been measured for SiC pixel detectors \cite{zhang2013}.

The radiation hardness of prototype SiC detectors constructed within the SiCILIA project \cite{tudisco2018} were tested by irradiating them with 60 MeV H$^+$ at a maximum dose of 2 kGy ($\approx 1.4 \cdot 10^{12}$ ions/cm$^2$). The reverse characteristics measured before and after the irradiation did not change significantly, and the leakage current measured up to -200 V remained low (10$^{-9}$- 10$^{-10}$ A) \cite{tudisco2019}. 
The performances of the device against the implantation of a heavy ion were investigated in Ref. \cite{raciti2010} and more recently in Ref. \cite{altana2023}, in which a Tandem beam of $^{16}$O at 25 MeV incident energy was used. The response of SiC and Si detectors were compared in terms of resolution and Charge Collection Efficiency (CCE) for increasing values of the ion fluence. The resolution of the Si detector deteriorated even at very low irradiation fluencies (10$^9$ ions/cm$^2$) and displayed very asymmetric peaks, whereas the SiC detector showed a small worsening of the resolution at larger fluencies (10$^{13}$ ions/cm$^2$) and the signal was always symmetric. 

\section{The prototype sensors}
\label{sec:prototype_sensors}

\begin{figure*}
\begin{center}
\includegraphics[width=\textwidth]{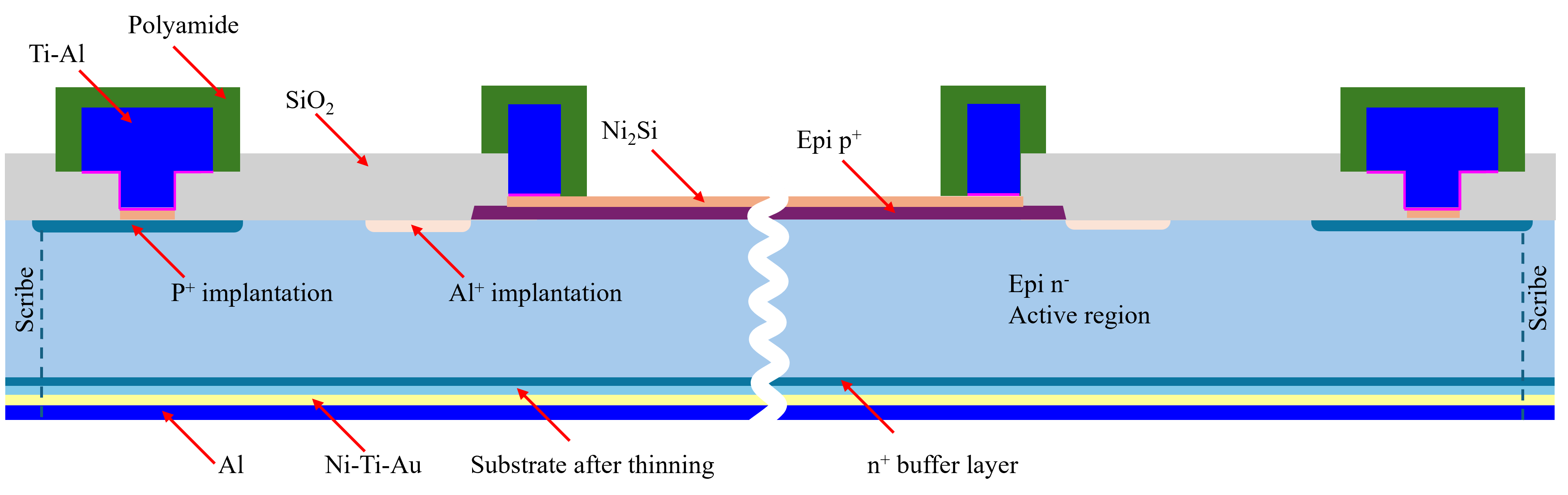}
\end{center}
\caption{A sectional view illustrating the internal structure of the prototype SiC device (not to scale).} 
\label{fig:sic_section}
\end{figure*}

New large area SiC sensors were produced from two epitaxial wafers as first prototypes to be used in NUMEN. The processing of these devices is similar to those developed within the SiCILIA project, described in Ref. \cite{tudisco2018}. The different layers present in the SiC section are sketched in Figure \ref{fig:sic_section}. 
They are produced starting from 100 $\mu$m n\textsuperscript{-} epi-layers grown on 350 $\mu$m SiC substrates. First, a double epitaxial layer, necessary to implement the p\textsuperscript{+}/n junction, is built growing a p\textsuperscript{+} layer 0.3–0.5 $\mu$m thick with an aluminum doping concentration of the order of 10\textsuperscript{18}-10\textsuperscript{19}/cm\textsuperscript{3} over the n\textsuperscript{-} epi-layer with a nitrogen doping concentration of $\approx$ 10\textsuperscript{13}/cm\textsuperscript{3}. Then, the detector and the edge structures areas are defined by multiple photolithographies steps \cite{tudisco2018}. The edge structures are built by the implantation of Al\textsuperscript{+} ions at the borders of the active region to reduce the electric field and P\textsuperscript{+} ions at the edge of the device creating n\textsuperscript{+} region that acts as a field stop. After these two implants, the dopants are activated by a high temperature process at 1650$\degree$ C. The deposition of an isolation oxide and the opening of the contacts with a further photolithographic process are then performed. The front of the device is metallized by a Nickel Silicide (Ni$_2$Si) deposition (thickness $\approx$ 100 nm) and annealed in order to form a good ohmic contact on the p\textsuperscript{+} and the n\textsuperscript{+} regions. A selective etch removes the unreacted Ni$_2$Si on the oxide and a thicker layer of Ti, and Al is deposited on the periphery of the detector for the bonding~\cite{tudisco2018}. Finally, front side passivation is performed by deposition of a thick polyamide layer, and the active area of the  detector is opened by a further lithography. On the back side of the device a mechanical thinning procedure is performed to reduce the total thickness of the device to $\approx$ 110 $\mu$m, which corresponds to a dead layer of $\approx$ 10 $\mu$m. Then, the ohmic contact is formed by an aluminum deposition ($\approx$ 1 $\mu$m). 

\begin{figure*}
\includegraphics[width=1\textwidth]{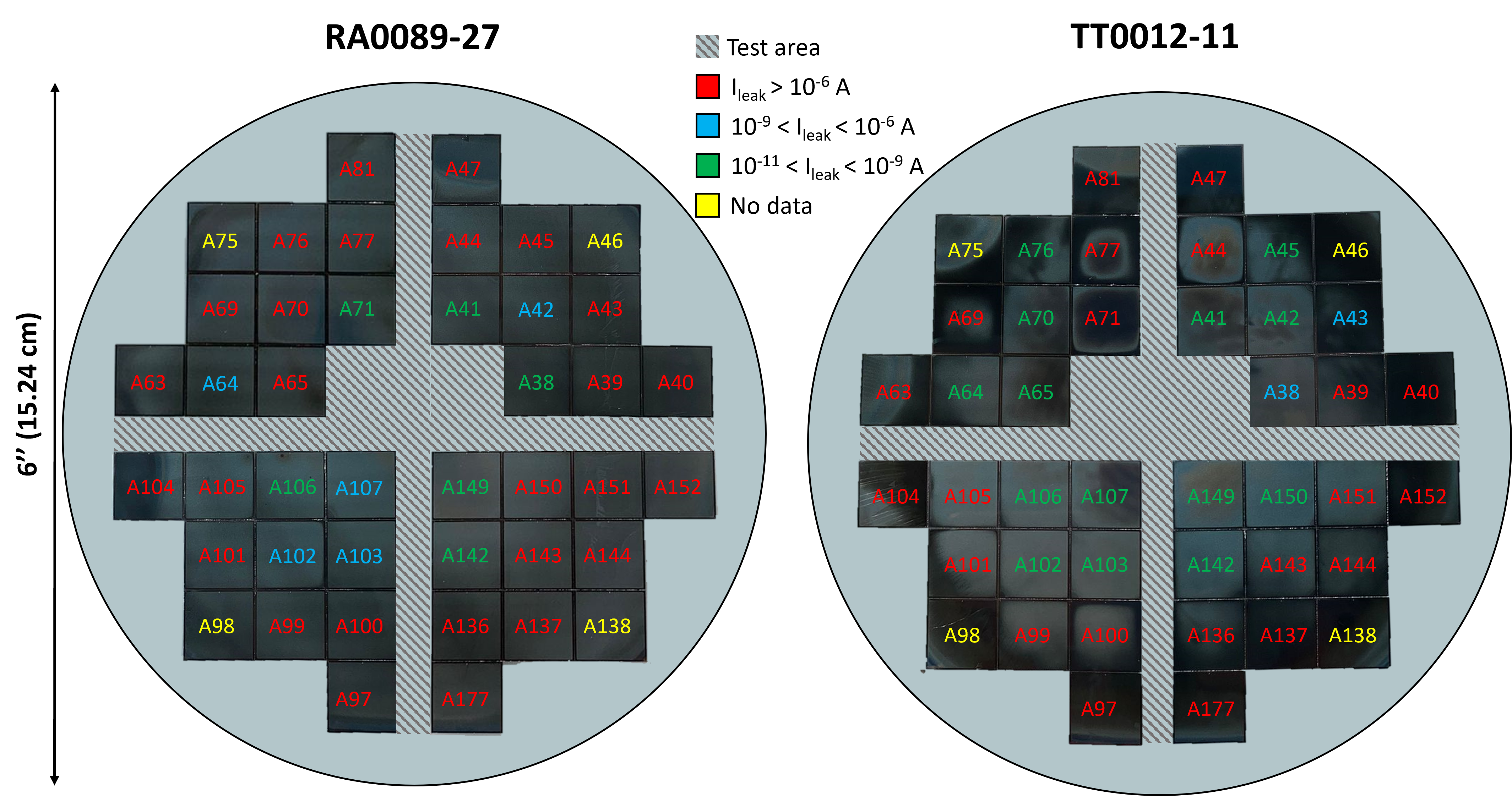}
\caption{SiC sensors belonging to the RA0089-27 and TT0012-11  wafers. The colored label indicate the class of measured inverse currents at maximum 200 V bias voltage, taken from the company datasheet.} 
\label{fig:wafers_picture}
\end{figure*}

The epitaxial wafers, coming from the same bulk material, used for producing the SiC detectors characterized in this work, are 6" diameter wafers named TT0012-11 and RA0089-27. The n\textsuperscript{-} epi-layer of TT0012-11 wafer was doped with the standard and well established concentration of $\approx$ 9 $\times$ 10$^{13}$ atoms/cm$^3$ corresponding to a Full Depletion Voltage (FDV) of $\approx$ 800 V for a $\approx$ 100 $\mu$m thick device. The RA0089-27 wafer n\textsuperscript{-} epi-layer was instead doped with lower concentration ($\approx$ 3 $\times$ 10$^{13}$ atoms/cm$^3$) as a case study for reducing the sensors FDV to a few hundred Volts. It is important to note that the reactor used to perform the epitaxy was pushed beyond its technical specifications in order to perform such low doping concentration. For this reason, we could expect worse performance from sensors produced from the RA0089-27 wafer. The device area is 15.4 $\times$ 15.4 mm$^2$. The total thickness of the sensors was measured by a micrometer to be 110 $\pm$ 1 $\mu$m. An edge structure $\approx$ 400 $\mu$m wide runs along the whole perimeter of the sensor. The frontside is equipped with a small pad (0.150 $\times$ 0.300 mm$^2$) for wire bonding, located in the center of one of its sides. The mask used to produce the devices from the 6" wafers made it possible to obtain 42 sensors from each one. The layout of the two wafers is shown in Figure \ref{fig:wafers_picture}. 

The two wafers were provided with datasheets from the company, in which the leakage current value for each sensor is reported. This first characterization performed by the company allowed for the subdivision of the 84 sensors in 3 main categories, based on the order of magnitude of the leakage current value: i) 10$^{-11}$ to 10$^{-9}$ A; ii) 10$^{-9}$ to 10$^{-6}$ A; iii) $>$ 10$^{-6}$ A. We emphasize that the maximum bias voltage applied in the company tests was 200 V, which was reached only for the first two sensor categories. The very large current value in the third category limited the bias voltages to lower values. It is expected that the good sensors, characterized by good energy and time resolution, belong to the first category, whereas sensors belonging to the third category can not be used as particle detectors. Regarding the second category, further investigation is needed to judge how these sensors behave in terms of energy and time resolution.
The inverse current values measured for the first category of good sensors correspond to a current density from 4 pA/cm$^2$ to 0.4 nA/cm$^2$, which are in good agreement with values reported in a recent paper on SiC microstrip detectors \cite{puglisi2019}.

The classification of the devices inside the two wafers is reported in Figure \ref{fig:wafers_picture} with different colors for the different inverse current categories. The four devices at the edge of the wafers, indicated in yellow, were not tested at the company. There is also a central region of the wafers (labelled as $\textit{test area}$ in Figure \ref{fig:wafers_picture}) devoted to a test pattern for process and technology development. Since the production of the two wafers was the first one for such large area SiC detectors, the test area was quite large. In the future it will be possible to reduce it, optimizing the area devoted to detector production. 

From a first inspection of the two wafers, we notice a difference in the production yield of good devices. Indeed, we find 14 good sensors in the TT0012-11 wafer, and only 6 good sensors in the RA0089-27 wafer (displayed in green in Figure \ref{fig:wafers_picture} and belonging to the first inverse current category). However, in the RA0089-27 wafer we find more sensors belonging to the second inverse current category with respect to the TT0012-11 wafer that need further inspections (indicated in blue in Figure \ref{fig:wafers_picture}). The production yield depends on the defect density of the SiC epitaxial films, which was measured for similar epitaxial growths in Ref. \cite{tudisco2018}. From that results it comes a production efficiency of $\approx$ 30\% for 2 cm$^2$ detectors. This value is in agreement with the one observed for the TT0012-11 wafer.

\section{IV and CV characterization}
\label{sec:iv_cv}

C-V scans were performed on several SiC sensors of the first two categories to characterize their FDV, and I-V measurements evaluated the leakage current at different bias voltages. The measurements were performed at INFN-Torino at the ‘Laboratorio Rivelatori Innovativi al Silicio’ using a Keysight B1505A Power Device Analyzer connected to a probe station. Two modules were used: i) an high voltage SMU B1513C, for device polarization and current measurements; ii) a multi frequency CMU B1520A, LRC meter for the capacitance measurements. The back of the sensor was biased with a positive voltage while a needle probe connected the pad on the top of the sensor to ground. The measurements were obtained using a RpCp representative model of the device under study. 

\begin{figure}
\begin{center}
\includegraphics[width=0.5\textwidth]{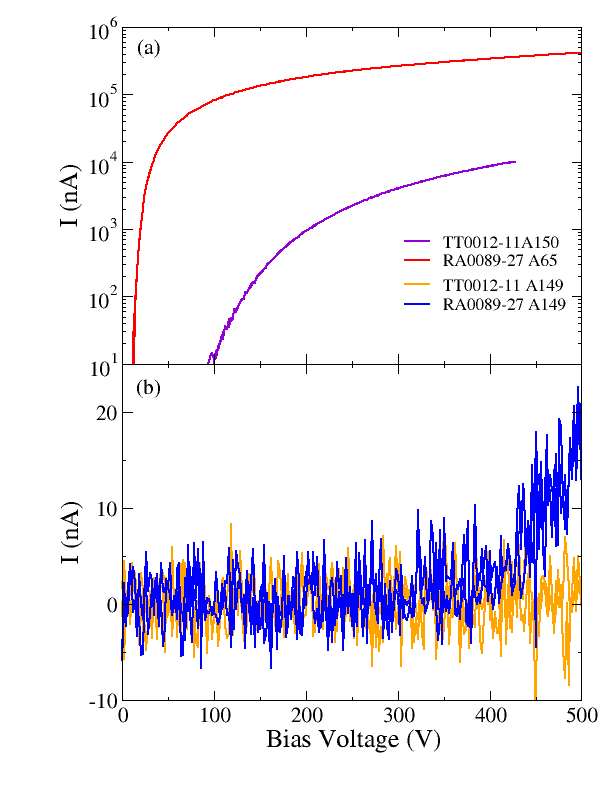}
\end{center}
\caption{I-V characteristics for some SiC sensors up to 500 V bias voltage. In panel (a) specific SiC sensors display high inverse currents, while in panel (b) other SiC sensors display the typical inverse current of a good detector.} 
\label{fig:IV}
\end{figure}

Concerning the I-V characteristics, typical trends of current measured for some sensors of both wafers are presented in Figure \ref{fig:IV}. In particular, the devices shown in Figure \ref{fig:IV}a are characterized by high values of the inverse current, from tens to hundreds of $\mu$A, making these sensors unusable as particle detection. In the case of A65 sensor of the RA0089-27 wafer, the current increases quickly already with a bias of some tens of Volt. The A150 sensor of the TT0012-11 wafer, which was classified as a good SiC from the company datasheets (see Figure \ref{fig:wafers_picture}), reaches 10 $\mu$A at $\approx$ 400 V and thus it is not suitable as particle detector. In Figure \ref{fig:IV}b, two typical inverse current trends for two good sensors of the two wafers are shown. The observed trend is quite stable up to 400 V. However, it is not possible to evaluate the absolute value of the inverse current of the two sensors with this measure, since the sensitivity of the instrument is $\approx$ 5 nA and we expect current values in the range 10$^{-11}$ to 10$^{-9}$ A from the datasheets of the company (see Section \ref{sec:prototype_sensors}). 
From the systematic investigation of the I-V characteristics for all the SiC sensors, we are able to test the original classification of the devices based on leakage currents. In most cases the original classification was confirmed. However, in only one case a device previously classified as good (the A150 of TT0012-11 wafer) was shown to be not working. On the other hand, in the RA0089-27 wafer we found three devices (A42, A64, and A102) belonging to the second category (indicated in blue in Figure \ref{fig:wafers_picture}) that can be considered good. 

\begin{figure}
\begin{center}
\includegraphics[width=0.5\textwidth]{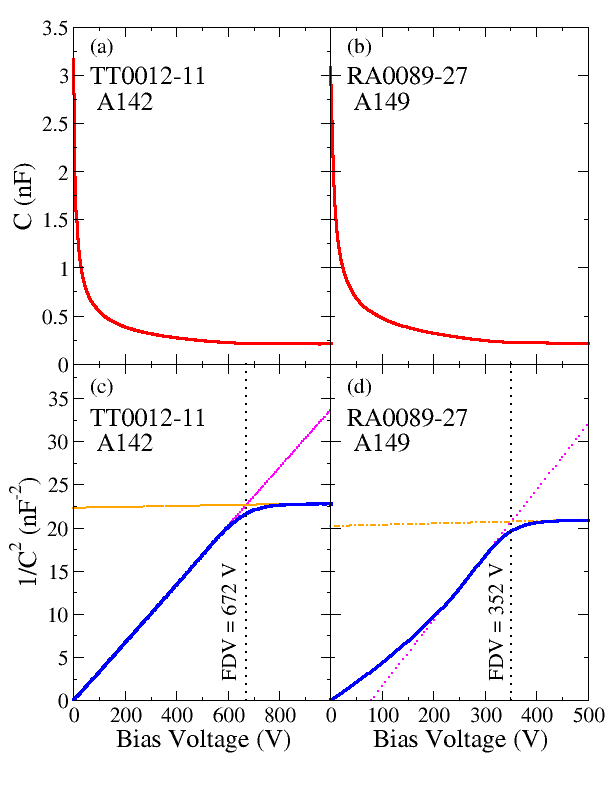}
\end{center}
\caption{Characterization for the TT0012-11 A142 and RA0089-27 A149 SiC detectors. Panels a) and b) C-V scan (red lines). Panels c) and d) 1/C$^2$ trend (blue lines) together with two linear fits of the two sections of the curve (orange and magenta dot-dashed lines). The FDV values are indicated as the dotted black lines.} 
\label{fig:A142CV}
\end{figure}

An example of C-V scan for a good sensor of the TT0012-11 wafer is shown in Figure \ref{fig:A142CV}a. The trend of 1/C$^2$, shown in Figure \ref{fig:A142CV}c, is characterized by the typical behavior: it is linear up to the knee after which it reaches a constant value. The FDV of the SiC device is extracted from the crossing point of the two straight lines resulting from the linear fits of the two sections of the 1/C$^2$ curve (as indicated in Figure \ref{fig:A142CV}c). The same representations are plotted for a good sensor of the RA0089-27 wafer in Figure \ref{fig:A142CV}b and \ref{fig:A142CV}d. We notice that the trend of the 1/C$^2$ curve does not follow a linear behaviour with increasing bias voltage, but it shows a flex, which would indicate that the doping concentration is not uniform into the sensor. Indeed, the reactor was not in its typical working conditions since the low doping concentration of the RA0089-27 wafer was at the limit of its technical specifications. This could result in a non-uniform concentration of the dopants inside the epitaxial layer.

\begin{figure}
\includegraphics[width=0.5\textwidth]{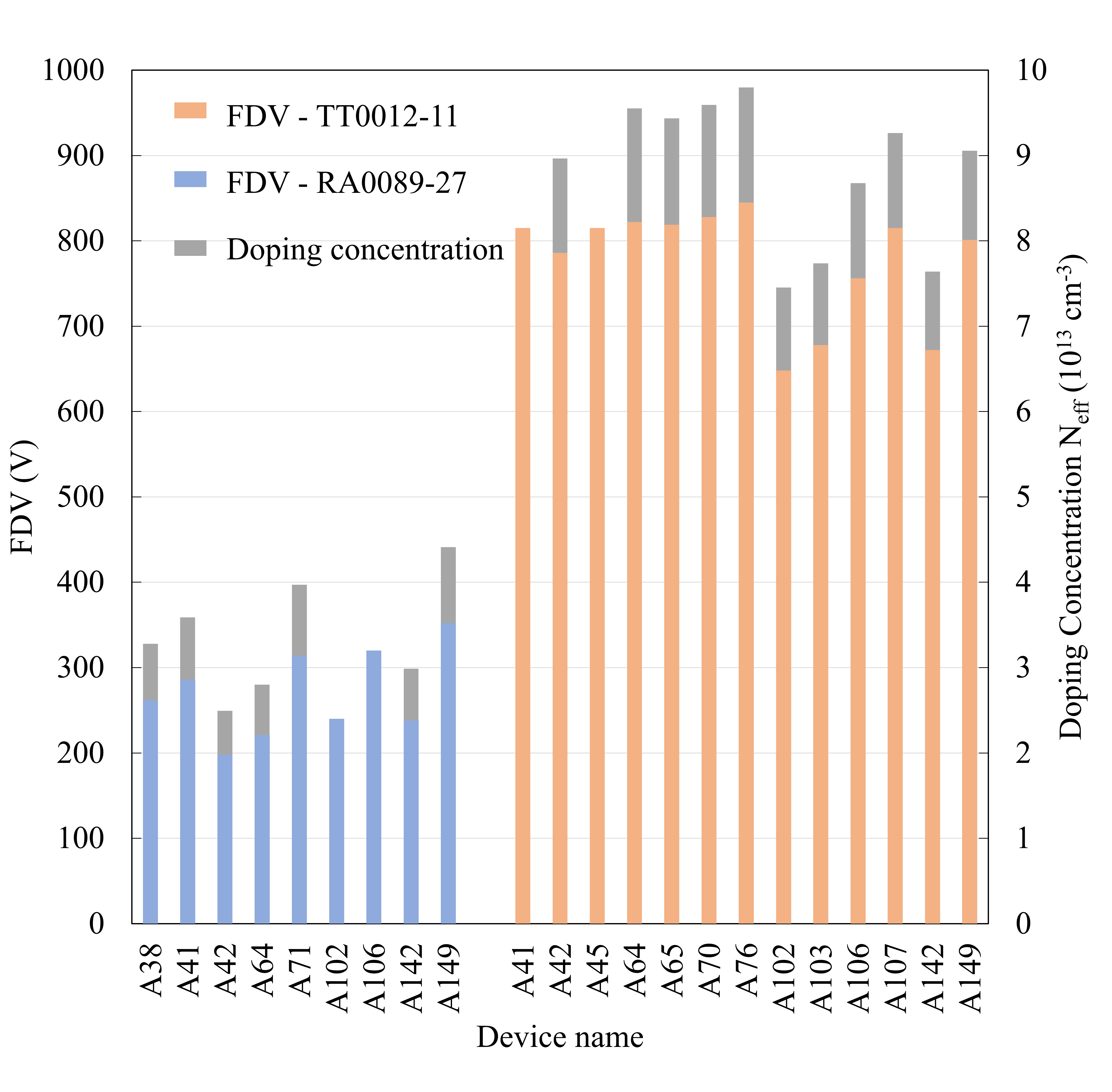}
\caption{Full depletion voltage (FDV) for the good SiC sensors belonging to the RA0089-27 and TT0012-11 wafers (blue and orange bars, respectively). The doping concentrations N$_{eff}$ for the same sensors is also reported (grey bars).} 
\label{fig:fvd_doping}
\end{figure}

The FDV values were extracted for all the good SiC sensors and are reported in Figure \ref{fig:fvd_doping}. The A102 and A106 devices of the RA0089-27 wafer and the A41 and A45 devices of the TT0012-11 wafer were not characterized by the probe station since they were connected to test boards for the $\alpha$-source measurements (see Section \ref{sec:alpha}). Their FDVs were determined by injecting a charge from a pulser on the back of the sensor that was biased to negative voltage on the front side. By measuring the height of the preamplifier output signal (V$_{out}$) as a function of the applied bias voltage (V$_{bias}$), we deduced the 1/C$^2$ trend being C(V$_{bias}$) $\propto$ V$_{out}$ \cite{schroder}.

It is evident from Figure \ref{fig:fvd_doping} that there are two sets of FDV values depending on the wafer of origin. The average FDVs for the RA0089-27 and TT0012-11 wafers are 270 V and 777 V, respectively. This reveals the different doping concentration in the two wafers. The mean absolute deviations (MAD) values of the FDVs are $\approx$ 43 V and $\approx$ 54 V for the RA0089-27 and TT0012-11 wafers, respectively. The large values of the MADs reflect a less than ideal uniformity of the devices inside the wafers, which can be an issue when managing a large number of detectors as in the case of the MAGNEX PID system that will be composed by 720 SiC sensors. 

\section{Doping profile}
\label{sec:doping}
The profile of the doping concentration $N$ as a function of the depth $W$ inside the semiconductor can be determined from the C-V characteristics. The most commonly used form was deduced by Hilibrand and Gold \cite{Hilibrand1960}, in which $N$ can be expressed as 

\begin{equation}
   N=\frac{2}{e\varepsilon A^2 \frac{d(1/C^2)}{dV}}
\label{eq:5}
\end{equation}  

\begin{figure}
\begin{center}
\includegraphics[width=0.5\textwidth]{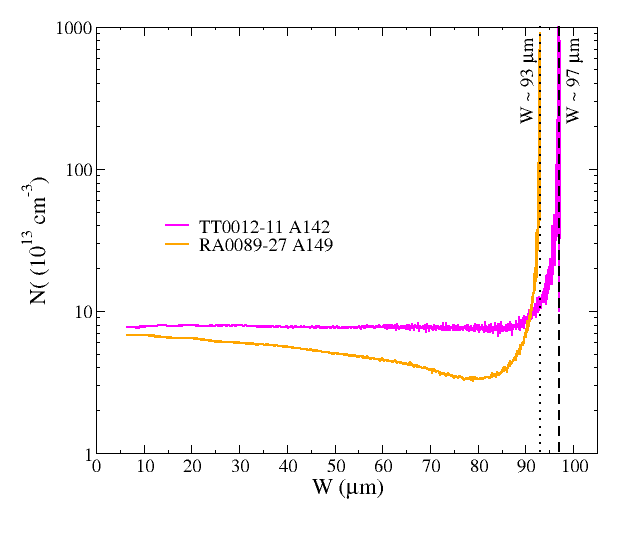}
\end{center}
\caption{Doping concentration profile vs. depth inside two SiC sensors. The A142 device of the TT0012-11 wafer and the A149 device of the RA0089-27 wafer are shown as the magenta and orange lines, respectively. The depletion depth is also indicated for the two curves.} 
\label{fig:doping_vs_depth}
\end{figure}

An example of the doping concentration profile for sensors belonging to the two wafers is shown in Figure \ref{fig:doping_vs_depth}. We adopted the value $\varepsilon_r$ = 9.7, according to Ref. \cite{tudisco2018}. The A142 SiC belonging to the TT0012-11 wafer exhibits the typical behaviour of the doping concentration as a function of the depth, which is quite constant up to 90 $\mu$m. The edge of the curve corresponds to the depletion depth that in this case is W $\approx$ 97 $\mu$m, close to the nominal 100 $\mu$m depth of the epitaxial layer. On the contrary, the A149 SiC of the RA0089-27 wafer shows a not uniform $N$ profile which decreases with depth. Moreover, the depletion depth is W $\approx$ 93 $\mu$m, significantly smaller than the nominal value. 

When the depletion region reaches the backward contact, the capacitance remains constant (see for example Figure \ref{fig:A142CV}a), this value is sometimes referred to as \textit{end capacitance} $C_{end}$. The voltage at which this situation occurrs is referred to as the full depletion voltage $V_{FDV}$. It is possible to define the effective doping concentration $N_{eff}$ in correspondence to $C_{end}$ and $V_{FDV}$ as found in Ref. \cite{petek2018} 

\begin{equation}
   |N_{eff}|=\frac{2V_{FDV}C_{end}^2}{e\varepsilon A^2}
\label{eq:6}
\end{equation} 

The effective doping concentrations were evaluated from the C-V characteristic for all the SiC devices analysed so far. Examples are represented in Figure \ref{fig:fvd_doping}. There is a clear correlation between the FDV and the doping concentration. In particular, for the TT0012-11 wafer, in which the average FDV is 777 V, the average doping concentration is  8.8 $\times$ 10$^{13}$ atoms/cm$^3$ (MAD = 0.7 $\times$ 10$^{13}$ atoms/cm$^3$), whereas for the RA0089-27, the doping concentrations are much smaller (average value 3.4 x 10$^{13}$ atoms/cm$^3$ and MAD = 0.5 $\times$ 10$^{13}$ atoms/cm$^3$), corresponding to an average FDV of 270 V. 

\section{Measurements with radioactive sources}
\label{sec:alpha}

Some of the detectors characterized in the previous sections were further investigated by using $\alpha$-particle source. In particular, the A41 and A45 sensors of the TT0012-11 wafer, as well as the A102 and A106 sensors of the RA0089-27 wafer were tested. The aim was to measure the energy resolution of the SiC sensors, and to estimate the depletion depth. A $^{228}$Th $\alpha$-decay source was chosen to perform such measurements, since it is characterized by a wide energy spectrum of the emitted alphas (E$_\alpha$ = 5 to 9 MeV), penetrating and probing the SiC device at different depths. Furthermore, the same source can be used to irradiate the device from the back, since the emitted alphas able to cross the $\approx$ 10 $\mu$m dead layer and release a measurable residual energy in the SiC active region. This approach was used to perform an energy-loss based measurement of the dead layer thickness to deduce the depletion depth at the FDV. 

\begin{figure}
\begin{center}
\includegraphics[width=0.25\textwidth]{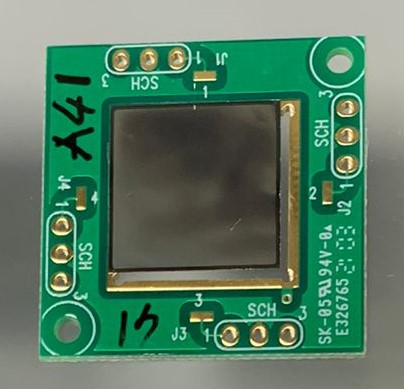}
\end{center}
\caption{Picture of the A41 sensor of the TT0012-11 wafer mounted within the PCB.} 
\label{fig:foto_sic}
\end{figure}

In a first experimental setup, an uncollimated 3.7 kBq $^{228}$Th source was placed $\approx$ 15 cm away the front of the SiC detector, inside a vacuum chamber (kept at a pressure of $\approx$ 10$^{-6}$ mbar). The source has a 100 $\mu$g/cm$^2$ thick Au window through which emitted $\alpha$ particles must pass. A circular entrance window of 10 mm diameter was defined by means of an aluminum plate mounted in front of the SiC sensor, in order to exclude the borders of the detector, which may be affected by partial charge collection effects. In this way, an angular aperture of $\pm 3^\circ$ was defined. The SiC device was placed within a suitably sized opening of a PCB, as shown in Figure \ref{fig:foto_sic}, and a conductive tape provided the electronic ground to the back side. The signal was collected from the front side that was wire bonded and placed at negative potential (typically between -900 to -200 V, according to the values reported in Figure \ref{fig:fvd_doping}). A 40 mV/MeV charge sensitive preamplifier, placed inside the vacuum chamber, was used to read-out the signal and to distribute the bias. The signal was shaped and amplified by the ORTEC 572 Amplifier, after which it was digitized by an ORTEC EASY-MCA ADC connected to the acquisition PC.

\begin{figure}
\begin{center}
\includegraphics[width=0.5\textwidth]{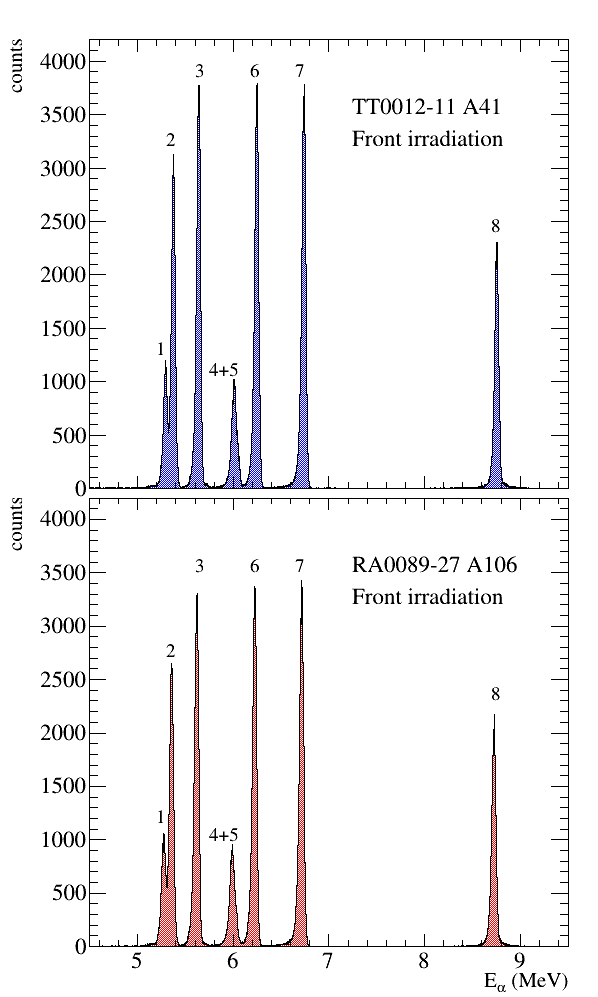}
\end{center}
\caption{Calibrated energy spectra measured with the $^{228}$Th $\alpha$-source irradiating the front side of the TT0012-11 A41 (top panel) and RA0089-27 A106 (bottom panel) sensors at their FDVs. Peak labels refers to the $\alpha$-decay energies listed in Table \ref{tab:th_source}.} 
\label{fig:spettri_front}
\end{figure}

Examples of $^{228}$Th $\alpha$-decay spectra for the A41 sensor of the TT0012-11 wafer and the A106 sensor of the RA0089-27 wafer at their FDVs are shown in Figure \ref{fig:spettri_front}. The ADC channel spectrum was calibrated in energy assuming the reference spectrum of the $^{228}$Th source, reported in Table \ref{tab:th_source}, taking into account the energy loss of the $\alpha$ particles crossing the Au window and the Ni$_2$Si deposit in the front side of the SiC sensor. The resulting residual energies of the $\alpha$ particles entering in the active region of the sensors are listed in Table \ref{tab:th_source}. For calibration purposes we considered only peaks labeled as 2, 3, 6, 7, and 8, since peaks 4 and 5 are very close to each other and peak 2 contains small contributions from $\alpha$ at 5448.6 and 5340.4 keV \cite{Artna1997_224,Artna1997_228}. The extracted energy resolution is $\approx$ 0.5$\%$ FWHM for all the SiC samples, extracted from the peak at 8784.86 keV. This result is much better than those reported in a recent review regarding state-of-the-art resolutions achieved with SiC detectors in $\alpha$ spectrometry \cite{mandal2023}. For the other peaks the extracted resolutions follow the $1/\sqrt{E}$ slope, which indicates that the electrical noise is negligible. 

\begin{table}
    \centering
    \caption{$\alpha$-decay spectrum of the $^{228}$Th source. The decaying nucleus of the $^{228}$Th decay chain, energy (E), branching ratio (BR) and residual energy (E$_{r}$) are listed.  E$_{r}$ is evaluated after crossing the Au window and the Ni$_2$Si layer. $\alpha$ energy values and BR from refs. \cite{Artna1992_212,Artna1997_216_220,Artna1997_224,Artna1997_228}}
\label{tab:th_source}
\small
    \begin{tabular}{cccccc}
    \hline
    Peak \# & Nucleus & E (keV) & BR (\%) & E$_{r}$ (keV)\\
    \hline
    1 & $^{228}$Th & 5340.36 & 27.2 & 5282\\
    2 & $^{228}$Th & 5423.15 & 72.2 & 5365 \\
    3 & $^{224}$Ra & 5685.37 & 94.9 & 5629\\
    4 & $^{212}$Bi & 6050.78 & 69.91 & 5996\\
    5 & $^{212}$Bi & 6089.88 & 27.12 & 6036\\
    6 & $^{220}$Rn & 6288.08 & 99.89 & 6235\\
    7 & $^{216}$Po & 6778.3 & 99.99  & 6728\\
    8 & $^{212}$Po & 8784.86 & 100.00 & 8741\\   
    \hline
    \end{tabular}
\end{table}

After the thinning procedure of the SiC substrate from the backside \cite{tudisco2018}, an inert layer of $\approx$ 10 $\mu$m remains beyond the epitaxial layer, which in principle is 100 $\mu$m thick. However, from the analysis of the doping profiles discussed above (see Figure \ref{fig:doping_vs_depth}), it seems that not only is the depletion depth reached at the FDV less than the expected 100 $\mu$m, but it is also different for the two wafers. Thus, in order to have a more accurate estimate of the depletion depth, the SiC detectors were irradiated with the $^{228}$Th source from the back side. In this way, the thickness of the dead layer present at the back can be extracted from the measured residual energy in the active region of the sensors. 

\begin{figure}
\begin{center}
\includegraphics[width=0.5\textwidth]{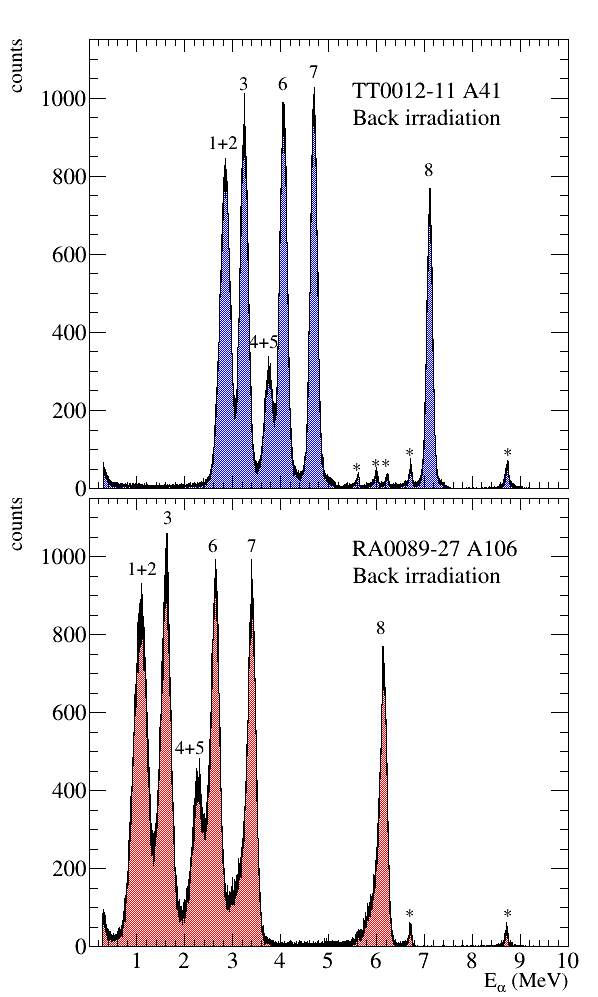}
\end{center}
\caption{Calibrated energy spectra measured with the $^{228}$Th $\alpha$-source irradiating the back side of the TT0012-11 A41 (top panel) and RA0089-27 A106 (bottom panel) sensors at their FDVs. Peak labels refers to the $\alpha$-decay energies listed in Table \ref{tab:th_source}. Peaks labeled with an asterisk correspond to the residual $^{228}$Th $\alpha$-source radioactivity from the decay of the $^{220}$Rn gaseous daughter nucleus.} 
\label{fig:back}
\end{figure}

The experimental setup was the same as used for the irradiation on the front side. We measured the $^{228}$Th $\alpha$ spectrum from the back of the SiC sensors at the FDV, and we calibrated it adopting the energy calibration deduced from the front side measurements. Examples of calibrated back spectra measured at the FDV for the A41 sensor of the TT0012-11 wafer and for the A106 sensor of the RA0089-27 wafer are shown in Figure \ref{fig:back}. Peaks from $\alpha$-source appear wider and at lower energy due to the inert SiC substrate crossed. We see in the spectra also other peaks, labeled with asterisks in Figure \ref{fig:back}, that correspond to the residual $^{228}$Th $\alpha$-source radioactivity from the decay of the $^{220}$Rn gaseous daughter nucleus that remains in the vacuum chamber for some days. The presence of these peaks confirms that the energy scale is correctly calibrated. 

We notice that peaks measured in the A106 sensors are at lower energy with respect to the A41 spectrum. From the residual energy values for each alpha peak, the corresponding SiC inert substrate thicknesses can be deduced from the energy loss tables \cite{Ziegler1985} of the LISE++ code \cite{Lise++}. The results are listed in Table \ref{tab:deadlayer}. The reported values correspond to the average and standard deviation of the thicknesses deduced from each alpha peak. The measured dead layers in the TT0012-11 sensors are compatible with the 10 $\mu$m dead layer expected from the thinning procedure. Instead, in the case of the RA0089-27 wafer the measured dead layer thickness are significantly larger. Since the total thickness of the devices is 110 $\pm$ 1 $\mu$m, measured by a micrometer, we can exclude that this is connected to the thinning procedure. The most probable scenario is that the sensors do not reach the full depletion depth of $\approx$ 100 $\mu$m. This finding confirms what was previously found from the analysis of the doping profile, reported in Section \ref{sec:doping}, and it is probably due to problems in the doping procedure of the RA0089-27 wafer.    

\begin{table}
    \centering
    \caption{Dead layer thickness and depletion depth at the FDV for the analysed SiC sensors.}
\label{tab:deadlayer}
    \small
    \begin{tabular}{cccc} 
    \hline
    Wafer & Sensor & Dead Layer  & Depletion depth  \\
        &  & ($\mu$m) &  ($\mu$m) \\
    
    \hline
    TT0012-11 & A41 & 9.6 $\pm$ 0.1 & 100.4 $\pm$ 0.1\\
    TT0012-11 & A45 & 10.8 $\pm$ 0.1 & 99.2 $\pm$ 0.1\\
    RA0089-27 & A106 & 14.6 $\pm$ 0.2 & 95.4 $\pm$ 0.2 \\
    RA0089-27 & A102 & 15.4 $\pm$ 0.2 & 94.6 $\pm$ 0.2 \\
    \hline
    \end{tabular}
\end{table}

\section{Conclusions}
\label{sec:conclusions}

First characterizations of state-of-the-art large area SiC detectors were performed in order to verify their performances in terms of the NUMEN project requirements. Sensors were produced from two different wafers, built from the same bulk material, doped with different doping concentration corresponding to different full depletion voltages. The values of the FDV and the doping profile were determined by measurements of the SiC I-V and C-V characteristics. Also the energy resolution and the depletion depth were deduced from $\alpha$-source irradiation tests. In general, the tested SiC sensors display a good energy resolution ($\approx$ 0.5 \% FWHM), which complies with the NUMEN requirements. Differences were found in the depletion depth for SiC belonging to the two wafers.

The observed spread in the FDV values ($\approx$ 7\% for the TT0012-11 wafer and $\approx$ 16\% for the RA0089-27 one) is quite large compared to the typical values for the standard Silicon detectors. However, this is the first time that p-n junction, large area ($\approx$ 2 cm$^2$), 100 $\mu$m thick epitaxial layer SiC devices are produced from 6" wafers, thus the uniformity is not known \textit{a priori}. The production of the devices studied in the present work was a pioneering one, aimed also to guide further R\&D activities. 

The TT0012-11 wafer was doped with $\approx$ 9 $\times$ 10$^{13}$ atoms/cm$^3$. Devices belonging to this wafer displayed good C-V characteristics and doping profile, and they reach a depletion depth of $\approx$ 100 $\mu$m, as expected. However, these sensors are characterized by quite high FDV, up to 950 V, which is challenging if we want to use more than 700 detectors in a low-pressure gas environment, as in the focal plane detector of the MAGNEX spectrometer, due to an high probability of discharges and electrical instabilities.

On the other hand, sensors belonging to the RA0089-27 wafer, doped with $\approx$ 3 $\times$ 10$^{13}$ atoms/cm$^3$, are characterized by a much lower FDV ($\approx$ 200-300 V), more suitable for the NUMEN purposes. However, from the analyses reported in this paper, their performance is worse, especially in terms of the doping profile and depletion depth. Moreover, looking at the production yield in the two wafers, we obtain about 31\% of good SiC detectors from the TT0012-11 wafer, in agreement with the expected one, and only 21\% from the RA0089-27 one. This could be connected to the worse doping profile obtained during epitaxy phase.

We can conclude that the reduction of the doping concentration to values around $\approx$ 3 $\times$ 10$^{13}$ atoms/cm$^3$ is mandatory for the purpose of the NUMEN project. It is crucial that the technology for such concentrations are improved in order to obtain good doping profiles and consequently good SiC detector performance as well as production yield. To this aim, a new reactor was developed, able to perform uniform layer growths at $\approx$ 3 $\times$ 10$^{13}$ atoms/cm$^3$. Motivated by the results shown in this paper, new wafers were recently doped by the new reactor and the obtained doping profiles seem to comply with the required standards of uniformity. They will be tested through the techniques reported in this paper, once the detector production is accomplished. 

In order to guide further R\&D activities on SiC sensors, investigations on the charge collection efficiency behaviour in the full active volume and near the edges of the detectors are of fundamental importance. This can be studied using 3D microscopic characterization techniques as the Ion beam-induced charge (IBIC) that utilizes focused MeV range accelerated ions, with beam spot as low as $\approx$ 1 $\mu$m, to probe charge transport \cite{Jaksic2022}.  

\section{Acknowledgements}
\label{}

This project received funding from the European Union "Next Generation EU" (PNRR M4 ‐ C2 – Inv. 1.1 - DD n. 104 del 02‐02‐2022 - PRIN 20227Z4HB8). The authors wish to thank Dr. Retief Neveling for helping the upgrade of the manuscript readability.



\bibliographystyle{elsarticle-num-names}

\bibliography{biblio.bib}


\end{document}